\documentclass[sigplan,screen,10pt]{acmart}

\renewcommand\footnotetextcopyrightpermission[1]{} 
\pdfoutput=1
\acmConference[Protecting Sensitive Tabular Data in Hybrid Clouds]{Protecting Sensitive Tabular Data in Hybrid Clouds}{2021}

\usepackage{amsmath}
\usepackage{graphicx}
\usepackage{multirow}
\usepackage{float}
\graphicspath{ {./images/} }

\AtBeginDocument{%
  \providecommand\BibTeX{{%
    \normalfont B\kern-0.5em{\scshape i\kern-0.25em b}\kern-0.8em\TeX}}}

\begin{document}

\title{Protecting Sensitive Tabular Data in Hybrid Clouds}


\author{Maya Anderson}
\affiliation{%
  \institution{IBM Research}
  \country{Israel}}
\email{mayaa@il.ibm.com}

\author{Gidon Gershinsky}
\affiliation{%
  \institution{IBM Research}
  \country{Israel}}
\email{gidon@il.ibm.com}

\author{Eliot Salant}
\affiliation{%
  \institution{IBM Research}
  \country{Israel}}
\email{salant@il.ibm.com}

\author{Salvador Garcia}
\affiliation{%
  \institution{Marina Salud Hospital}
  \country{Spain}}
\email{SALVADOR.GARCIA@marinasalud.es}


\begin{abstract}
Regulated industries, such as Healthcare and Finance, are starting to move parts of their data and workloads to the public cloud. However, they are still reluctant to trust the public cloud with their most sensitive records, and hence leave them in their premises, leveraging the hybrid cloud architecture. We address the security and performance challenges of big data analytics using a hybrid cloud in a real-life use case from a hospital.  In this use case, the hospital collects sensitive patient data and wants to run analytics on it in order to lower antibiotics resistance, a significant challenge in healthcare.
We show that it is possible to run large-scale analytics on data that is securely stored in the public cloud encrypted using Apache Parquet Modular Encryption (PME), without significant performance losses even if the secret encryption keys are stored on-premises. PME is a standard mechanism for data encryption and key management, not specific to any public cloud, and therefore helps prevent vendor lock-in.  It also provides privacy and integrity guarantees, and enables granular access control to the data. We also present an innovation in PME for lowering the performance hit incurred by calls to the Key Management Service. Our solution therefore enables protecting large amounts of sensitive data in hybrid clouds and still allows to efficiently gain valuable insights from it.
\end{abstract}



\maketitle

\section{Introduction}
Hybrid clouds are composed of an on-premises IT infrastructure, fully controlled by a company or a public organization, and of a public cloud infrastructure shared by users from many companies and organizations. The public cloud provides isolation between different users and supports strong security guarantees for data, so one cloud tenant is unable to view or modify the data or other resources of another cloud tenant. Still, businesses are often wary about placing their sensitive parts of information in a shared public cloud. The hybrid cloud architecture addresses this challenge by allowing organizations to keep their "crown jewels" and controls (such as encryption keys and user verification) inside the business premises, while using the public cloud for storage of sensitive data that was encrypted on premises, and for data processing.
Despite the benefits presented by the hybrid architecture, it still faces a number of technical challenges related to performance of data processing pipelines that crunch large amounts of sensitive information. The compute, storage and security components of such pipelines are spread across the various public and private domains, with a tightly controlled connection between the on-premises and public cloud infrastructures. The data processing on analytic workloads have a degraded performance if the data is encrypted.
In this work, we address the security and performance challenges of big data analytics in hybrid cloud use cases by leveraging the new Apache Parquet standard for columnar data encryption - Parquet Modular Encryption (PME). As a part of the Apache Parquet community, we are contributing to the design and implementation of PME \cite{ggstrata19} , including its key management layer which is optimized for highly distributed architectures such as hybrid clouds. In this paper, we demonstrate these technologies via a real-life use case from a hospital that leverages both its on-premises IT infrastructure and public cloud services. Hybrid cloud architecture provides an additional benefit for such users, since it allows saving a significant investment in building and maintaining an on-premises IT infrastructure, along with the expensive IT and security in-house specialists required to plan and deploy such systems. Much of this burden is offloaded to the public cloud that is fully staffed with IT and security experts, who provide efficient support to a large number of cloud users and therefore provide advanced but affordable services to the customers. The customers, such as the hospitals, can focus on their main area of professional expertise - and also can keep the critical security components (such as encryption key managers) inside their infrastructure, so they are never deployed in a public cloud.  This approach allows having stronger security at a lower price, since the commodity tasks are performed by cloud providers' teams of experts in public clouds, while the critical data access components, such as encryption keys and user authorization, are deployed in-house.

\section{Parquet Modular Encryption in Hybrid Cloud}

Parquet modular encryption has been a part of the Apache Parquet file format standard since version 2.7 of the format specification \cite{cryptospec}. Parquet files are a popular option for storing and processing large amounts of analytic data due to the encoding, compression, column filtering and row group filtering capabilities of its format. Instead of fetching and crunching a full data set from the storage, Parquet allows the analytic frameworks to perform much of selection of the required data in storage, and fetches only small data subsets directly relevant for the analytic queries. 

\subsection{Envelope Encryption}

As part of the Apache Parquet community, we have contributed to the design and development of an encryption key management layer for Parquet Modular Encryption \cite{kmt}. This design adopts the practice of \textit{envelope encryption} where a \textit{data encryption key} (DEK) is itself encrypted, or \textit{wrapped}, with another key (such as a \textit{master encryption key}, MEK) and stored near the encrypted data. MEKs are managed by a Key Management Service (KMS), that handles both storage and access control of the master keys. These keys are also called \textit{root keys} and are stored internally in the KMS and used for key wrapping. Parquet Modular Encryption key tools create DEKs using a random number generator, a single DEK per Parquet column chunk, and encrypt (wrap) them using MEKs. Two wrapping modes were devised:
\begin{enumerate}
    \item The common mode – single wrapping: In single-wrapping mode, the random DEK is encrypted directly with MEK. The drawback of this mode is that there is interaction with KMS for every DEK, to wrap it with the MEK.
    \item The novel mode – double wrapping: In order to minimize interactions with a KMS server, a double-wrapping mode can be used. Each \textit{data encryption key} (DEK) is encrypted with a \textit{key encryption key} (KEK), that in turn is encrypted with a \textit{master encryption key} (MEK). In a writer process, a random KEK is generated for each MEK  and is cached in the process for a limited time, configurable by the user. This allows the process to perform an interaction with a KMS server only once for each MEK, in order to wrap its KEK. DEK wrapping is performed locally, and does not involve interaction with a KMS server. Therefore, many files and columns that use the same MEK can be encrypted with many DEKs without having to interact with the KMS for DEK wrapping. The time limit is imposed on the KEK cache, so that authorization by the KMS is not compromised. After a preset amount of time, PME will have to call the KMS in order to wrap/unwrap a KEK, and if access to the master key was in the meantime revoked, the call will fail. This closely fits the operation model of the hospital, which batches writes to the data store, typically completing all writes within ten minutes - which is a reasonable period for fitting within the timeout period for KEK use.
\end{enumerate}

\section{Use Case Details}

Marina Salud is a private company managing the health of the population within the Marina Alta area of Spain. The size of this area is comparable to the city of Geneva and is home to approximately 200,000 year-round inhabitants. The population includes a large number of retirees making the effective management of chronic conditions a high priority.

Marina Salud has a network of 34 primary care facilities, 2 ambulatory centers and a 206 bed hospital.     
In 2015 Marina Salud Hospital established an Antimicrobial Stewardship Program (ASP), called PROA by its acronym in Spanish. The main objective of this program is to ensure that broad spectrum antibiotics are used only when needed, based on medical evidence, as the overuse of this kind of medication is increasing bacterial multi-resistance, reducing the effect of the medication.

As its name implies, the program is composed of a group of medical experts in antibiotics who continually monitor the prescriptions prescribed by the rest of the physicians. When an incorrect prescription is detected, a recommendation is provided which the prescribing physician is free to accept  or not. As such, the recommendations from the PROA team serve as training for the rest of physicians with less expertise in this area.

Monitoring is performed for various cases that may indicate an improper prescription or the need to change an existing prescription and is accomplished by querying millions of medical records, combining diagnoses, lab results and prescription data. The time span across which the queries operate is limited in order to allow the processing  to complete in an acceptable time (less than a minute).

In particular, the PROA team looks for:
\begin{enumerate}
    \item Patients with a SEPSIS alert who have been prescribed a broad spectrum antibiotics in the last 30 days
    \item Patients without a SEPSIS alert prescribed a broad spectrum antibiotics in the last 3 days
    \item Patients diagnosed with bacteremia in the last 15 days
    \item Patients diagnosed with bacterial multi-resistance in the last 15 days
    \item Patients with a quick Sequential Organ Failure Assessment (qSOFA) alert in the last 15 days
\end{enumerate}

 The queries are carried out on a clustered relational database with two nodes.
 
 Typically in the hospital, medical records collected during the day are batched, and then uploaded to long-term data store in one shot.  In Marina Salud Hospital, this means that approximately 1000 records for 300 patients are uploaded over approximately a ten minute period once a day.
 
For our work, we wanted to test the performance of queries on large amounts of medical data which are representative of queries being done today in the Marina Salud Hospital for antimicrobial-stewardship.  While we wanted to use only synthetically generated data to both avoid privacy issues and to facilitate the acquisition of large amounts of data, we also wanted the data to closely resemble real world medical records, both semantically and from a content point of view.   

Our  representative query looks for  the improper prescribing of the antibiotic, Amoxicillin, by querying Parquet-encrypted Electronic Medical Records (EMRs) to get a list of patients who were prescribed Amoxicillin but without having observed symptoms of  either 'Acute bronchitis' , ‘Otitis media’, ‘Acute bacterial sinusitis’ or ‘Sinusitis’ within a time frame starting with the initial encounter date, and ending two days after the encounter end date.  It is interesting to note that in Marina Salud Hospital, with millions of patient medical records, this query is limited to only searching three days back for performance reasons - otherwise it would simply take too long.

Synthetic medical records in FHIR format were generated using the widely utilized Synthea (Synthetic Patient Population Simulation) generator \cite{synthea}.  The generation of the EMRs was slightly biased to include a higher incidence of records based around ear infections and bronchitis, through the application of modules corresponding to these ailments.

\section{HL7 Fast Healthcare Interoperability Resources (FHIR)}
FHIR is the new, emerging interoperability standard for the exchange of healthcare information in the healthcare ecosystem. FIHR specifies a model for a wide range of data objects in this ecosystem  - known as resources - as well as a specification for the exchange of these resources, typically through a RESTful interface. 

We implemented the storage of FHIR resources as encrypted Parquet files by adding a special Interceptor to the IBM FHIR Server \cite{ibmfhir}.  In this configuration, generated FHIR resources are sent in JSON over REST to the FHIR server, where each REST resource has its own URI suffix.

The FHIR server deserializes the received REST call, and upon identifying a request to persist data, triggers a hook to supplied Interceptor code, which in turn uses Apache Spark \cite{spark} to write the resources out in encrypted Parquet format.  Encrypted files are stored in subdirectories according to their resource type.

\begin{figure}[htp]
    \centering
    \includegraphics[width=\columnwidth]{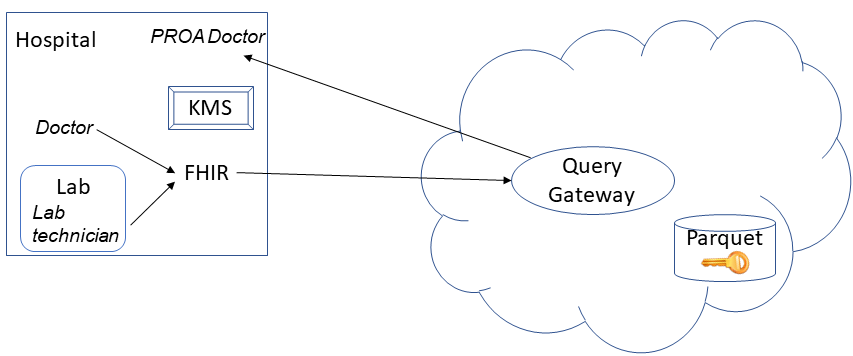}
    \caption{Use case diagram}
    \label{fig:hospitals}
\end{figure}

We also enabled SQL querying of stored encrypted Parquet resources.  In our solution, SQL queries are sent out as REST calls to our Query Gateway server, which uses Spark SQL with PME support to create virtual tables, where each table represents a resource, and consequently executes the passed query, returning the result to the caller. Figure \ref{fig:hospitals} shows that the Query Gateway is deployed in the cloud, and so is the storage where the encrypted Parquet resources reside. On the other hand, the Key Management Service is deployed on the hospital premises (Private Cloud).

\section{Evaluation}

In order to analyze the overhead of protecting sensitive data in Hybrid Clouds we measured the run-time of the query described in the Use Case Details section. We used the Hashicop Vault \cite{vault} key management service to store the master keys. The data used was synthetic medical record of four sizes: 10,000, 20,000 40,000 and 60,000 patients, stored in 4 Parquet tables in a Cloud Object Store. The respective sizes of the folder with the Parquet files are: 39MB, 119MB, 230MB and 341MB. Since Apache Spark does lazy evaluation, we persist the result of the query into another Parquet, so the measured time includes both the query time and the time for persisting the result. We rune the query on a 3-node Spark cluster.

Before the experiments Parquet files are encrypted using Parquet Modular Encryption. Each Parquet file has 3 to 5 sensitive columns encrypted using one master key, with the other columns encrypted using another master key. In addition, the file footers are encrypted using a third master key. All master keys are created per table; different tables have different master keys and hence potentially a different set of eligible readers, having a different set of access permissions. Spark is located remotely from the  Key Management Service to represent the situation where the encryption master keys are saved on customer premises in Hashicorp Vault, but the Spark analytics engine and the data are in the Cloud.

\begin{figure}[htp]
    \centering
    \includegraphics[width=\columnwidth]{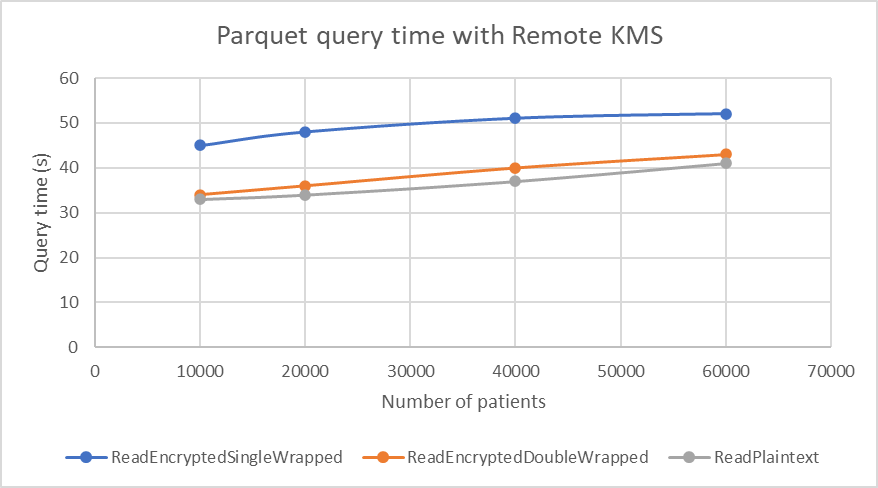}
    \caption{Query times - remote KMS}
    \label{fig:remotekms}
\end{figure}

\begin{table}[htp]
    \caption{Decryption/Encryption Overhead - remote KMS}
    \label{table:remotekms}  
    \begin{tabular}{|l|l|l|l|l|}
    \toprule
        Size & 10k & 20k & 40k & 60k \\ 
        \midrule
        \multirow{2}{7em}{ReadEncrypted Single Wrapped} & 45 & 48 & 51 & 52 \\
        &  &  &  &  \\ \hline
        \multirow{2}{7em}{ReadEncrypted Double Wrapped} & 34 & 36 & 40 & 43 \\
        &  &  &  &  \\ \hline
        ReadPlaintext & 33 & 34 & 37 & 41 \\ \hline
        \multirow{2}{10em}{Single Wrapped Encryption Overhead} & 36\% & 41\% & 38\% & 27\% \\
        &  &  &  &  \\ \hline
        \multirow{2}{10em}{Double Wrapped Encryption Overhead} & 3\% & 6\% & 8\% & 5\% \\ 
        &  &  &  &  \\ 
        \bottomrule
    \end{tabular}
\end{table}

Comparing the run time of querying the plaintext data with that of querying the encrypted data in Figure \ref{fig:remotekms} and in Table \ref{table:remotekms}, we can see that the overhead of decrypting the source files and then encrypting the result files using keys managed by a KMS that is remote from Spark using single-wrapped keys is 27\% to 41\%. However, if double-wrapped keys are used then this overhead drops to 3\% - 8\%.

\begin{figure}[htp]
    \centering
    \includegraphics[width=\columnwidth]{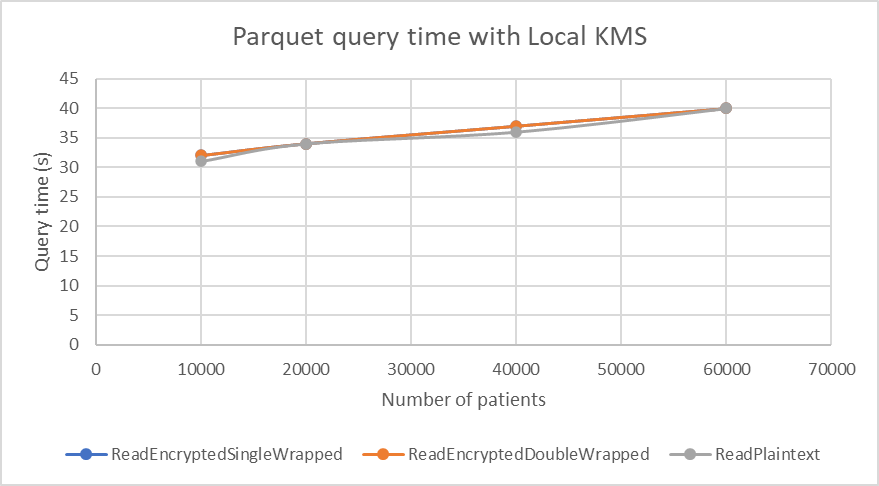}
    \caption{Query times - local KMS}
    \label{fig:localkms}
\end{figure}

In order to demonstrate that the reason for the decryption and encryption overhead is that the KMS is remote, we also ran the same suite of tests with the same data on the same Spark where KMS is located near Spark. In this case we see in Figure \ref{fig:localkms} and in Table \ref{table:localkms} that the overhead becomes 3\% to 6\% in single wrapping, and  0\% to 8\% in double wrapping, as was expected  based on the performance analysis of Parquet modular encryption with Java 11 \cite{perfblog}.

\begin{table}[htp]
    \caption{Decryption/Encryption Overhead - local KMS}
    \label{table:localkms}  
    \begin{tabular}{|l|l|l|l|l|}
    \toprule
        Size & 10k & 20k & 40k & 60k \\ 
        \midrule
        \multirow{2}{7em}{ReadEncrypted Single Wrapped} & 32 & 34 & 37 & 40 \\
        &  &  &  &  \\ \hline
        \multirow{2}{7em}{ReadEncrypted Double Wrapped} & 32 & 34 & 37 & 40 \\
        &  &  &  &  \\ \hline
        ReadPlaintext & 31 & 34 & 36 & 40 \\ \hline
        \multirow{2}{10em}{Single Wrapped Encryption Overhead} & 3\% & 0\% & 3\% & 0\% \\
        &  &  &  &  \\ \hline
        \multirow{2}{10em}{Double Wrapped Encryption Overhead} & 3\% & 0\% & 3\% & 0\% \\ 
        &  &  &  &  \\ 
        \bottomrule
    \end{tabular}
\end{table}

The round-trip of wrapping or unwrapping a key using a master key in Vault, when Spark is remote from the KMS, is around 470ms.  We have 4 input tables, with 3 master keys each, and the result is one output table with 3 master keys, so the expected upper limit on the overhead of wrapping/unwrapping keys when using double wrapping and hence using every master key just once:
\[4 \cdot 3 \cdot 470 \textrm{[wrapping]} + 3 \cdot 470 \textrm{[unwrapping]} =  7050ms = 7 s\]
This estimation is an upper limit, since it assumes that the keys are retrieved serially.
Therefore, the actual overhead of 3 seconds or less in Table \ref{table:remotekms} is indeed within these bounds.

\section{Related Work}

Colombo et al. \cite{colombo} describe a framework for Data Protection as a Service (DPaaS) to cloud computing users in the Multi-Cloud Environment. In addition to supporting the basic data encryption capability, this DPaaS framework allows data owners to define fine-grained access control policies to protect their data. Data protected by an access control policy are automatically encrypted and access is granted to user/applications according with the policy. This work, though also presenting client-side cloud-independent encryption, is different in that it protects data on a coarser granularity – no per-column encryption, but on a file or block level. It trusts storage for anti-tampering, as opposed to our usage of GCM for integrity protection, and it mandates an agent process deployed in a VM.

Xu and Zhao \cite{xu} present a framework for privacy-aware computing on Hybrid Clouds with Mixed sensitivity data. They model data sensitivity using a set of tagging mechanisms – from file level  to line level, temporal and spatial, and they can process data dynamically generated on-the-fly. However, their approach to the Hybrid cloud is to guarantee data privacy by segregating the sensitive data from the rest, and processing the sensitive data on the private cloud only.

There are other articles about such secure frameworks, but in our approach no specialized framework is needed, we use an open source Apache Spark software that is very commonly used for large-scale analytics.

Envelope encryption is treated briefly in the NIST Special Publication 800-57, Recommendation for Key Management \cite{nist}, under the name "key wrapping". 

\section{Conclusions}

We have shown how using Parquet Modular Encryption (PME) in a hybrid cloud environment can protect sensitive hospital data while allowing running analytics on the large-scale data. Double key wrapping innovation in PME significantly reduces the amount of time required to allow for the decryption of data during analysis. This was demonstrated on a real-life Healthcare use case from a hospital whose purpose is reducing antibiotics resistance.

\begin{acks}
The CyberKit4SME and ProTego projects leading to this publication have received funding from the European Union’s Horizon 2020 research and innovation programme under grant agreements No 883188 and 826284.
\end{acks}

\bibliographystyle{ACM-Reference-Format}
\bibliography{sample-base}

\end{document}